\documentclass[%
 aip,
 jmp,%
 amsmath,amssymb,
 reprint,%
]{revtex4-1}

\usepackage{graphicx}
\usepackage{epstopdf}
\usepackage{bm}
\usepackage{float}
\usepackage[normalem]{ulem}
\usepackage{dcolumn}
\usepackage{bm}
\usepackage{color}
\usepackage{footnote}

\usepackage{color, soul}

\begin{document}

\preprint{APS/123-QED}

\title{A HBAR-oscillator-based 4.596~GHz frequency source: Application to a coherent population trapping Cs vapor cell atomic clock}
\author{Thomas Daugey, Jean-Michel Friedt, Gilles Martin, 
and Rodolphe Boudot}

\affiliation{FEMTO-ST, CNRS, UFC, 26 chemin de l'Epitaphe 25030 Besan\c{c}on Cedex, France
}

\date{\today}

\begin{abstract}
This article reports on the design and characterization of a high-overtone bulk acoustic wave resonator (HBAR)-oscillator-based 4.596~GHz frequency source. A 2.298~GHz signal,
generated by an oscillator constructed around a thermally-controlled two-port AlN-sapphire HBAR resonator with a Q-factor of 24000 at 68$^{\circ}$C, is frequency multiplied by 2
to 4.596~GHz, half of the Cs atom clock frequency. The temperature coefficient of frequency (TCF) of the HBAR is measured to be $-23$~ppm/$^{\circ}$C at 2.298~GHz.
The measured phase noise of the 4.596~GHz source is $-105$~dBrad$^2$/Hz at 1~kHz offset and $-150$~dBrad$^2$/Hz at 100~kHz offset. The 4.596~GHz output signal is used as a local oscillator (LO) in
a laboratory-prototype Cs microcell-based coherent population trapping (CPT) atomic clock. The signal is stabilized onto the atomic transition frequency by tuning finely
a voltage-controlled phase shifter (VCPS) implemented in the 2.298~GHz HBAR-oscillator loop, preventing the need for a high-power-consuming direct digital synthesis (DDS).
The short-term fractional frequency stability of the free-running oscillator is 1.8 $\times$ 10$^{-9}$ at one second integration time. In locked regime, the latter is improved in a preliminary proof-of-concept experiment at the level of
6.6 $\times$ 10$^{-11}~\tau^{-1/2}$ up to a few seconds and found to be limited by the signal-to-noise ratio of the detected CPT resonance.
\end{abstract}

\maketitle
\clearpage
\section{Introduction}\label{sec:introduction}
Over the last decade, outstanding progress in microelectromechanical systems (MEMS) technologies
and semi-conductor lasers, combined to coherent population trapping (CPT) physics \cite{Alzetta:1976, Arimondo:1996}, has allowed the development of high-performance miniature
atomic clocks (MAC) that combine a volume of about 15 cm$^3$, a total power consumption lower than 150 mW and a fractional frequency stability lower than 10$^{-11}$ at 1 hour and 1 day integration time \cite{Knappe:APL:2004, Knappe:Elsevier:2010, Lutwak:PTTI:2007}. Such frequency references, now commercially-available \cite{microsemi}, can provide the base for a number of mobile and embedded applications including network synchronization, new-generation mobile telecommunication systems, satellite-based navigation systems on-earth receivers, secure banking data transfer or military and avionic systems. In a Cs atom-based MAC, a 4.596~GHz local oscillator (LO), ultimately frequency-stabilized onto the atomic clock transition frequency, is required to drive a vertical-cavity-surface-emitting laser (VCSEL).

This local oscillator needs to satisfy numerous stringent requirements. It must be integrated onto a reduced surface footprint ($<$ 1 cm$^2$), consume a negligible maximum power of about 50 mW for battery-powered applications, deliver an output microwave power tunable between $-6$ and a few dBm \cite{Brannon:MTTSymp:2005, Romisch:IFCS:2006} to drive the VCSEL laser and be compatible with low-cost mass production. The LO output frequency must be tunable with a high resolution so that the relative frequency error remains significantly smaller than the targeted clock stability. In general, a frequency resolution at the mHz level is desired to ensure a frequency stability below 10$^{-12}$. Additionally, the output 4.596~GHz signal must be modulated in frequency with a rate and a depth of about 1~kHz in order to be used in a lock-in amplifier configuration. Moreover, phase noise is critical because the short term frequency stability of a passive atomic frequency standard can be degraded by the LO phase noise through an intermodulation effect \cite{Audoin:TIM:1991, Dick:PTTI:1987}. The
fractional frequency stability limitation $\sigma_{y_{LO}} (\tau)$ of the atomic clock due to this aliasing effect can be approximated by:
\begin{equation}
\sigma_{y_{LO}} (\tau=1\mbox{~s}) \sim \left(\frac{f_m}{\nu_0}\right) \sqrt{S_{\varphi}(2f_m)}
\label{eq:dick}
\end{equation}
where $S_{\varphi}(2f_m)$ is the power spectral density (PSD) of the LO phase fluctuations in the free-running regime at Fourier frequency $f = 2 f_m$ and
$\tau$ the averaging time. According to this relation, assuming a LO modulation frequency $f_m$ of 1~kHz and targeting $\sigma_{y_{LO}}$ $<$ 2 $\times$ 10$^{-11}$ at 1 s integration time, we calculate that $S_{\varphi} (2 f_m=2\mbox{~kHz})$ must be lower than $-77$~dBrad$^2$/Hz.

The most common technological approach for the development of a LO in MAC application, using discrete components, consists of a frequency synthesizer using a LC voltage-controlled oscillator (VCO) phase-locked to a 10 MHz quartz oscillator through a fractional-N phase-locked loop (PLL). In \cite{Lutwak:PTTI:2007}, a 4.596-GHz synthesizer was built reaching a power consumption of 40 mW and giving a clock short-term stability of $5 \times 10^{-11}~\tau^{-1/2}$, limited by the RF synthesizer phase noise. In Ref. \cite{a}, a 12-mW single-chip 4.596~GHz frequency synthesizer ASIC implemented in a 130-nm RF CMOS process was developed for a Cs MAC, demonstrating a clock stability of $5 \times 10^{-11}$ at 1 s, limited by the signal-to-noise ratio of the detected CPT signal. However, in such systems, the frequency multiplication degrades
the phase noise and can consume up to 50 \% of the MAC total power budget \cite{Lutwak:PTTI:2007}. In an another way, Brannon et al. developed
for MAC applications microwave oscillators based on quarter-wavelength ceramic-filled coaxial resonators. Nevertheless, these sources exhibited modest phase noise performances of $-62$~dBrad$^2$/Hz and $-94$~dBrad$^2$/Hz at 1~kHz and 10~kHz from the carrier respectively \cite{Brannon:MTTSymp:2005}.

A promising alternative solution for MAC applications is the development of microwave MEMS oscillators based on bulk acoustic wave resonators (BAW). These resonators exhibit an extremely small size and power consumption, very high Q$\times$f products up to 10$^{14}$ \cite{Lakin:UFFC:2005} and high-potential for wafer-level fabrication. In this domain, High-overtone Bulk Acoustic Resonators (HBAR) have emerged as a valuable tradeoff between thin-film bulk acoustic resonators (FBAR) -- reaching the microwave frequency range but requiring a thin and fragile membrane to be fabricated -- and Surface Acoustic Wave (SAW) resonators that benefit from a mature technology level, advanced
design tools, simpler manufacturing steps compared to thin-film-based devices but whose operating frequency is limited by lithography resolution.
Numerous HBAR-based oscillators have been developed (see for example \cite{Lakin:MTT:1993, Pang:MTTsymp:2005, Yu:UFFC:2007, Gachon:Ultrason:2007, Yu:UFFC:2009}). However, because of the difficulty to tune the LO frequency to the exact atomic transition frequency, we found a unique example in the literature of a HBAR-based oscillator frequency-locked to a CPT resonance. In 2009, Yu et al. demonstrated a HBAR-based 3.6~GHz Pierce oscillator with a phase noise of $-77$~dBrad$^2$/Hz at 1~kHz
Fourier frequency, a power consumption of 3.2 mW, a frequency stability of 1.5 $\times$ 10$^{-9}$ at 1 second \cite{Yu:UFFC:2009} in free-running regime and claimed to be improved at the level of 1 $\times$ 10$^{-10}$ when locked to a cm-scale Rb vapor cell.  Note that in this reference, the output signal of the HBAR oscillator was mixed with a RF signal from an external synthesizer, highlighting the difficulty to tune the output frequency to the exact atomic transition frequency. Moreover, a vapor cell with cm-scale dimensions was used in the CPT experiment. Also, no figures were reported on the CPT clock experiment in this article.

This article reports on an original double-port AlN/Sapphire 2.298~GHz HBAR-oscillator-based 4.596~GHz frequency synthesizer used as a local oscillator in a Cs vapor-microcell based CPT atomic clock. While not fully-integrated, the system architecture presented in this work is found to be well suitable for MAC applications with high potential for miniaturization and low power consumption. Dedicated design rules, "juggling" with the resonator dimensions, materials, temperature sensitivity and multi-mode spectrum specificity of HBAR, are provided for the resonator to be best suited to this application and to ensure the output frequency to be resonant with Cs atom at 4.596~GHz. Section \ref{sec:HBAR} describes the resonator. Section \ref{sec:HBARosc} describes the HBAR-based 2.298~GHz oscillator and 4.596~GHz frequency synthesizer. Section \ref{sec:cpt} details the CPT experiment using the HBAR-based frequency source.

\section{The HBAR resonator}\label{sec:HBAR}

An HBAR is obtained by stacking a thin piezoelectric film over a low-acoustic loss substrate. 
The thin piezoelectric film generates an acoustic wave by converting an incoming 
exciting electrical signal, while the thick substrate confines the acoustic energy, allowing for 
high quality factor. This coupled resonant structure yields a complex multimode spectrum 
in which a broad envelope, defined by the thin film resonance frequency and its overtones, 
modulates the admittance real part maximum of a comb of modes equally spaced by a frequency 
splitting defined by the inverse of the substrate thickness.

\begin{figure}[h!tb]
\includegraphics[width=\linewidth]{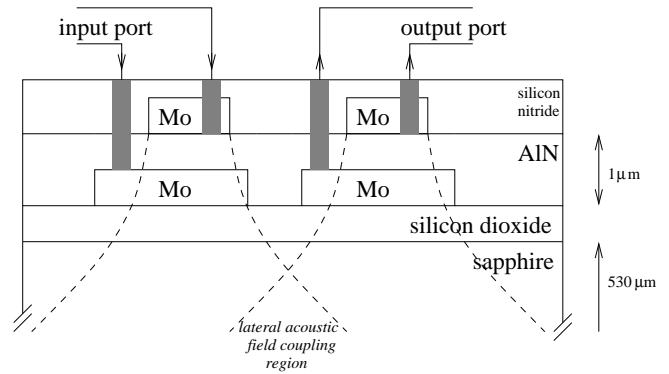}
\caption{Schematic cross section of the 2-port laterally coupled HBAR. The filled rectangles
(grey) indicate Ni-Au filled vias, while the dashed line hints at the acoustic field
distribution for defining the region of lateral acoustic field coupling between the two
transducers.}
\label{fig:hbar_sch}
\end{figure}

Figure \ref{fig:hbar_sch} shows the architecture of the HBAR resonator.
The approach, inspired by \cite{Chretien:IUS:2013, reinhardt2013ultra}, is to build two single-port HBAR resonators 
close enough one another in order to promote acoustic coupling via evanescent waves by proper selection of material 
combinations. The bottom electrode is shared by the two resonators and is accessed through dedicated vias connected to ground. 
Typical losses are in the 15 to 20~dB range, with a rejection better than 15~dB. This double-port resonator configuration is of great interest to simplify the development of an oscillator. Unlike the negative resistance Clapp, Pierce or Colpitts oscillators which use the anti-resonance of single-port resonators, dual-port resonator oscillators allow a rational approach of Barkhausen conditions. Additionally, the double-port configuration is less sensitive to parasitic electrical elements than standard oscillator architectures (Colpitts, Clapp or Pierce).

Five degrees of freedom are available when designing an HBAR: piezoelectric thin film and
low acoustic loss substrate material properties, thicknesses, and electrode geometry.
The piezoelectric substrate is selected with a high coupling coefficient for each mode to be
well defined, since each piezoelectric layer overtone spreads its coupling coefficient over
all the modes of the comb included in each envelope. Here a thin film of aluminum nitride (AlN)
is deposited under vacuum. The low-acoustic loss substrate defines each mode $Q$-factor: the lower the
losses, the better the quality factor. Here, sapphire is selected for its excellent properties
in the microwave frequency range \cite{braginsky}.

The resonance frequency $f_p$ of the piezoelectric film fundamental mode is fixed by the ratio $c_p/t_p$ with $c_p$ 
the longitudinal wave velocity in the piezo electric substrate and $t_p$ the piezoelectric film thickness. 
Here, considering the addition of Mo electrodes a longitudinal wave velocity $c_p$ of 11 270 m/s in AlN and 
a reached thickness $t_p$ of about 1~$\mu$m, the frequency of the fundamental mode is $f_p=c_p/t_p\simeq 1.25$~GHz. 
Consequently, the second overtone is used to reach the frequency of $2 f_p\simeq 2.5$~GHz, close to the quarter of 
the Cs clock frequency (2.298~GHz). The choice of the substrate thickness, that defines the spacing between 
adjacent modes, results from several key points. On the one hand, the thinner the substrate, the wider the 
spacing and the easier the oscillator design for selecting the targeted mode, but the more fragile the resonator. 
Obviously, without any other freedom degrees such as the resonator temperature, an excellent manufacturing 
resolution on the substrate thickness would be required in order to tune finely the resonator frequency to 
the exact atomic transition frequency. In a miniature Cs vapor cell clock, the CPT resonance linewidth ranges 
from a few kHz to 20 kHz. This order of magnitude determines the required resolution $\delta f$ on the HBAR 
resonance position accuracy. Let us assume a targeted operation frequency of about 4~GHz and a mode spacing of 
about 10~MHz as justified later. The overtone number would be $N=400$. Assuming that the piezoelectric layer 
acts as an acoustic energy pump and that the mode frequency is governed by the substrate, the uncertainty $\delta t_s$ 
on the substrate thickness $t_s$ is given by: $$\delta t_s=\frac{N\times c_s}{2\times f_0^2}\times
\delta f$$ with $c_s\simeq 10^4$~m/s the acoustic wave velocity in the substrate.
We calculate that reaching a frequency uncertainty $\delta f$ below 20~kHz requires a control of the substrate 
thickness such as $\delta t_s\simeq 2.5$~nm, that is not achievable with our technology.

Therefore we decided to adopt a two-step tuning strategy of the HBAR-based oscillator to ensure the output signal 
to be resonant with Cs atoms. A first coarse frequency tuning is reached by exploiting the frequency-temperature dependence 
of the HBAR. 
A second fine 
frequency tuning over a few tens of kHz is achieved by tuning the bias point of a voltage controlled phase shifter 
(VCPS) in the oscillator loop (see section \ref{sec:HBARosc}). 
The latter tuning capability must be used cautiously since the multi-mode nature of the resonator induces frequency
jumps if the VCPS bias point is changed so that an adjacent mode is favored.

\begin{figure}[h!tb]
\begin{center}
\includegraphics[width=0.85\linewidth]{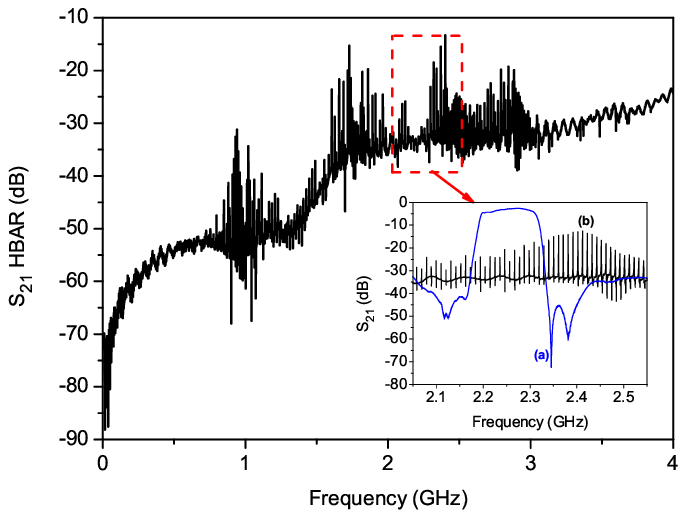} \\
(a) \\
\includegraphics[width=0.85\linewidth]{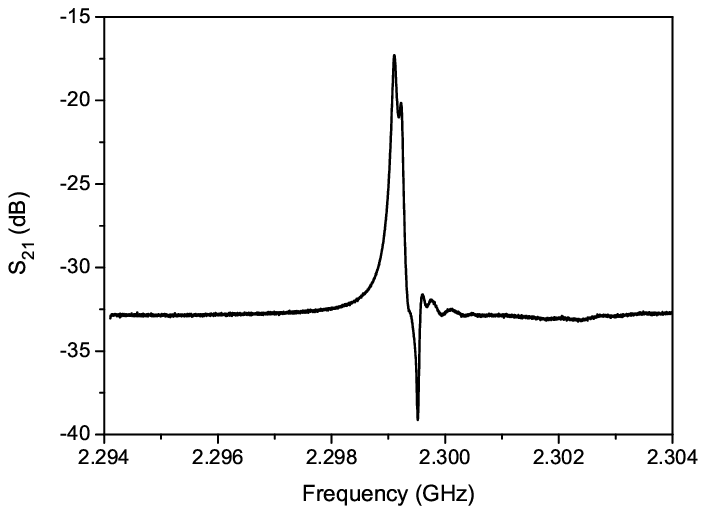} \\
(b)
\end{center}
\caption{
(a): S$_{21}$ parameter (magnitude) of the HBAR resonator over a large span of 4~GHz. The inset shows a zoom from 2.05 to 2.55~GHz. In the inset, the blue line (a) plots the transfer function of the SAW filter used to select a reduced number of acoustic modes. The other curve shows the response of the HBAR. (b): Selected resonance mode of the HBAR resonator at about 2.298~GHz. The HBAR temperature is here 55$^{\circ}$C. The Q-factor is measured to be 24 200.}
\label{fig:res-hbar-lspan}
\end{figure}

Figure \ref{fig:res-hbar-lspan} (a) displays for a wide span of 4~GHz the measured S$_{21}$ parameter (magnitude in dB) of the HBAR-resonator. The multi-modal structure of the HBAR is clearly shown. Zooming in the 2.298~GHz region, modes that can be used for the present Cs vapor cell atomic clock application are highlighted. A 100-MHz wide bandpass surface acoustic wave (SAW) filter (Golledge TA0700A) of central frequency 2.25~GHz is used for primary modal selection. In the filter bandwidth, about ten modes of the resonator response are selected and can be used for oscillation. Figure \ref{fig:res-hbar-lspan} (b) shows the spectrum (S$_{21}$ parameter, magnitude in dB) of the mode selected for the oscillator loop. Resonance and anti-resonance are well observed. Due to the double-port configuration, the resonance mode is found to be split into two neighbour distinct modes.  For a resonator temperature of 55$^{\circ}$C, insertion losses are measured to be about 17 dB. The resonance quality-factor $Q$, defined as $Q = \frac{\nu_0}{\Delta \nu}$ where $\nu_0$ is the resonance frequency and $\Delta \nu$ the 3-dB bandwidth, is measured to be 24 200. 

\begin{figure}[t]
\centering
\includegraphics[width=\linewidth]{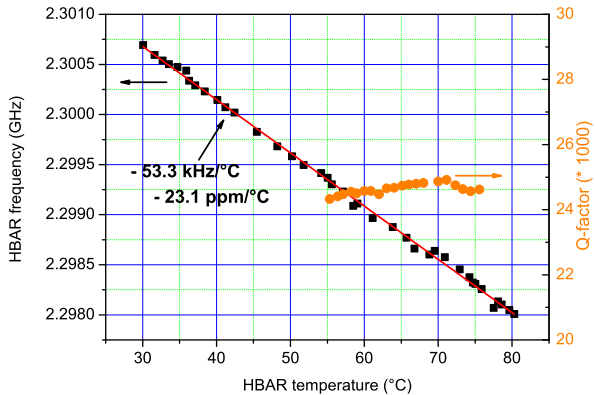}
\caption{HBAR frequency (filled squares, left axis) and loaded Q-factor (filled circles, right axis) versus the resonator temperature. Frequency measurement data are fitted by a linear function (solid line) with a slope of $-53.3$~kHz/$^{\circ}$C.}
\label{fig:f-Q-T}
\end{figure}

Figure \ref{fig:f-Q-T} shows the evolution of the HBAR frequency and Q-factor versus the resonator temperature in the 30 $-$ 80$^{\circ}$C range. The temperature coefficient of frequency (TCF) of the resonator is measured to be $-23.1$~ppm/$^\circ$C, i. e $-53.3$~kHz/$^\circ$C. We measured that the resonator Q-factor does not change significantly in this temperature range.

The temperature sensitivity of $-53.3$~kHz/$^\circ$C helps to define here the substrate thickness. Considering that we wish to keep the resonator below a temperature of 220$^{\circ}$C, the temperature span from room temperature to the maximum operating temperature is about 200~$^\circ$C. Thus, the tuning capability
of the frequency of each mode is 10.66~MHz, and defines the mode spacing. In our case, the mode spacing defined by a sapphire substrate of thickness $t_s=530$~$\mu$m
\cite{chretien:tel-01056972} propagating a wave at a velocity $c_s=$11 000~m/s \cite{ruppel2000advances} is $c_s/t_s/2=10$~MHz. Consequently, the HBAR is always able to bring one of its mode at 2.298~GHz, quarter of the Cs microwave transition. 


\section{The HBAR-oscillator}\label{sec:HBARosc}
\begin{figure}[h!tb]
\centering
\includegraphics[width=\linewidth]{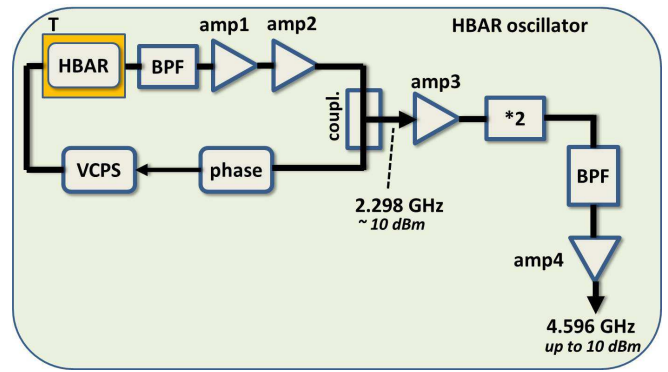}
\caption{
Architecture of the HBAR-based 2.298~GHz oscillator loop and frequency multiplication chain to generate an output 4.596~GHz signal. Amps 1, 2, 3 and 4 are microwave amplifiers. BPFs are band-pass filters. VCPS: voltage-controlled phase shifter, phase: variable phase shifter, coupl.: coupler, $\times$ 2: frequency doubler. References of the components are given in the text.}
\label{fig:osc-hbar}
\end{figure}

Figure \ref{fig:osc-hbar} shows the oscillator loop constructed around the two-port HBAR resonator. The resonator is stabilized at 68$^{\circ}$C with a high-precision temperature controller described in \cite{Boudot:RSI:2005}. At this temperature, insertions losses of the resonator are about 17 dB and the Q-factor is 24 200. 

\begin{figure}[h!tb]
\centering
\includegraphics[width=\linewidth]{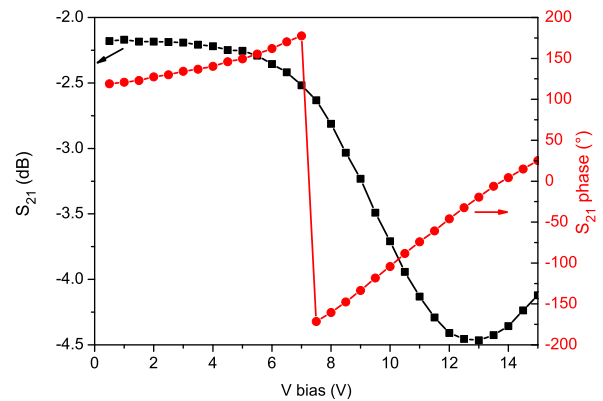}
\caption{Response (S$_{21}$ parameter) of the VCPS versus the control bias voltage. Filled squares, left axis: magnitude in dB. Filled circles, right axis: phase ($^{\circ}$).}
\label{fig:dct}
\end{figure}

Two sustaining amplifiers (Mini-circuits ZX60-3011+ and ZX60-3018G-S+), with a total gain of 30 dB, are used to compensate losses of the circuit after selecting the 10
modes around 2.25~GHz with the Golledge TA0700A bandpass filter. A phase shifter is implemented to tune the correct phase needed to meet the Barkhausen condition required for oscillation. A voltage controlled phase shifter (VCPS) (Minicircuits JSPHS2484) is implemented in the loop. This VCPS is used for fine tuning of the oscillator loop phase and consequently of the local oscillator output frequency. Figure \ref{fig:dct} exhibits the S$_{21}$ parameter (magnitude in dB and phase) versus the control bias voltage. The VCPS is operated around a bias voltage set point
that allows to obtain a high voltage-to-phase (in $^{\circ}$/V) dependence. Simultaneously, it should be preferred to operate the VCPS at a bias voltage where insertion losses of the VCPS are small to sustain oscillation and where the induced residual amplitude modulation is reduced. In our experiment, the chosen set point was close to 6 V. Around this value, the tuning voltage-frequency sensitivity was measured to be 8~kHz/V. 

\begin{figure}[h!tb]
\centering
\includegraphics[width=\linewidth]{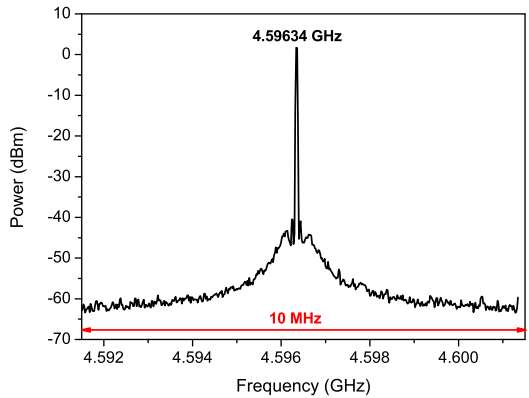}
\caption{Output spectrum of the 4.596~GHz signal over a 10 MHz span measured with a microwave spectrum analyzer. RBW, VBW: 30~kHz.}
\label{fig:spectre-4600MHz}
\end{figure}

The useful 2.298~GHz output signal of the HBAR-based oscillator loop is extracted using a microwave coupler (Mini-circuits MC ZABDC20-252-S+). This signal is then amplified with a microwave amplifier (Minicircuits ZX60-8008E-S+) and frequency-multiplied to 4.596~GHz with a low noise frequency doubler (Mini-circuits ZX90-2-36S+). The 4.596~GHz signal is bandpass-filtered with a 50-MHz bandwidth bandpass filter and amplified again to a power up to 10~dBm with an amplifier (Minicircuits ZX60-8000E-S+). A variable attenuator is used at the output to adjust the microwave power that drives the VCSEL laser. Figure \ref{fig:spectre-4600MHz} shows the spectrum of the 4.596~GHz output signal measured with a spectrum analyzer.

\begin{figure}[h!tb]
\centering
\includegraphics[width=\linewidth]{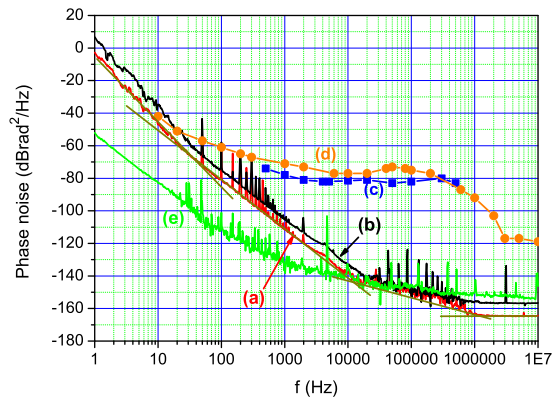}
\caption{Absolute phase noise performances. (a): HBAR-based oscillator 2.298~GHz signal, (b): HBAR-based synthesizer output 4.596~GHz signal, (c): 4.596~GHz synthesizer ASIC developed in \cite{a}, (d): 4.596~GHz synthesizer used in the commercially-available CSAC SA-45 from Microsemi, (e): state-of-the-art 100 MHz OCXO ideally reported to 4.596~GHz \cite{Francois:RSI:2015}. Solid lines at 2.298~GHz are used to extract $b_i$ coefficients of the phase noise spectrum power-law such as $S_{\varphi}(f) = \sum_{i=0}^{-4} b_if^i$.}
\label{fig:pnoise-hbar}
\end{figure}

Figure \ref{fig:pnoise-hbar} shows the absolute phase noise of the HBAR-based oscillator 2.298~GHz signal and the output 4.596~GHz signal respectively. For comparison, these phase noise performances are compared to those of a low-power consumption (12 mW) 4.596~GHz frequency synthesizer ASIC developed in \cite{a} for MAC applications, and to those of a low power 4.596~GHz frequency synthesizer developed in \cite{Lutwak:PTTI:2007}. Also, phase noise performances of a state-of-the-art 100 MHz oven-controlled quartz-crystal oscillator (Pascall OCXOF-E-100) ideally reported to 4.596~GHz are given \cite{Francois:RSI:2015}. The phase noise spectrum of the free-running 2.298~GHz signal is given in dBrad$^2$/Hz by the power law $S_{\varphi}(f) = \sum_{i=0}^{-4} b_if^i$ with $b_0=-165$, $b_{-1}=-102$, $b_{-3}=-20$ and $b_{-4}=-6$. Intersection between the $f^{-1}$ and the $f^{-3}$ slopes at 10 -- 20~kHz does not fit correctly with the expected Leeson frequency $f_L = \nu_0 / 2 Q_L \sim$ 47~kHz \cite{Leeson}, with $Q_L$ the HBAR loaded Q-factor and $\nu_0$ the HBAR frequency. We suspect that the expected Leeson frequency value is hidden by a non-negligible contribution of the LO amplitude noise in the 10~kHz -- 1~MHz region \cite{Rubiola}. Due to the multiplication factor by 2, the absolute phase noise of the 4.596~GHz is found to be 6 dB higher, except for the phase noise floor that is limited by the residual noise of the by-2 multiplication chain to $-157$~dBrad$^2$/Hz. The phase noise of the 4.596~GHz signal is measured at the level of $-105$~dBrad$^2$/Hz for an offset frequency $f=1$~kHz. Compared to the 4.596~GHz synthesizer ASIC developed in \cite{a}, phase noise performances of the present HBAR-based source are 26 dB and 67 dB better for $f=1$~kHz and $f=100$~kHz respectively. Compared to the 100-MHz OCXO-based synthesizer, phase noise performances of the HBAR-based source are found to be 25 dB worse for 
$f=1$~kHz and 5~dB better for $f=1$~MHz respectively. This demonstrates the high potential of HBARs for the development of ultra-low phase noise microwave sources, with high potential for low power consumption and miniaturization.

For information, according to Eq. (\ref{eq:dick}), using the present HBAR-based frequency source and assuming $f_m=1$~kHz, the Dick effect contribution to the clock fractional frequency stability $\sigma_{y_{LO}}$ ($\tau=1$~s) is rejected at the level of 4.3 $\times$ 10$^{-13}$. This remarkable value is below the ultimate atomic clock quantum noise limit for microcell-based miniature CPT atomic clocks \cite{ShahKitching:2010}.

\section{CPT atomic clock experiment}\label{sec:cpt}
\subsection{Description}

In a Cs CPT MAC, Cs atoms confined in a miniature vapor cell interact in a so-called lambda-scheme with a resonant bi-chromatic optical field generated by a laser system. When the frequency difference between both optical lines exactly equals the atomic ground state hyperfine splitting, atoms are trapped through a destructive quantum interference process into a coherent superposition of two long-lived ground state hyperfine levels. In this dark state, atoms are under ideal conditions fully decoupled from the excited state, resulting in an increase of the atomic vapor transparency. The resulting CPT resonance, whose linewidth is ultimately limited by the atom-light interaction time, can be used as a narrow frequency discriminator for the development of an atomic clock.


\begin{figure}[h!tb]
\centering
\includegraphics[width=\linewidth]{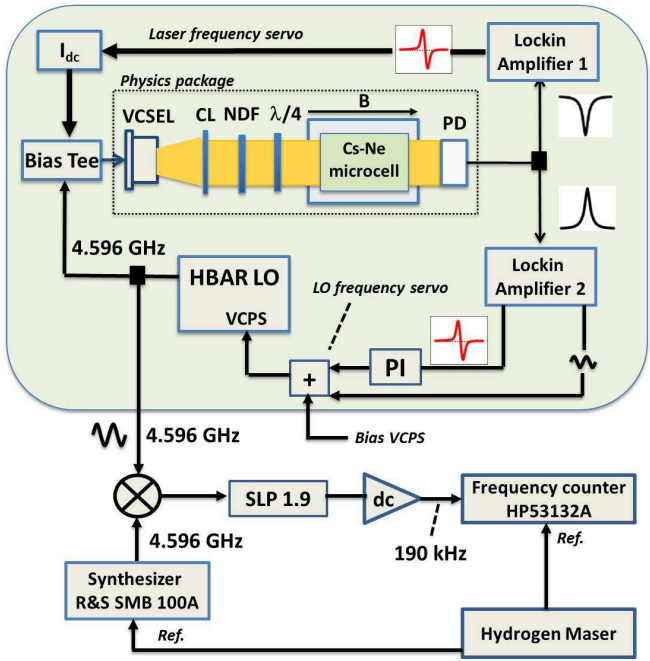}
\caption{Experimental setup of the CPT clock. A physics package contains the VCSEL laser, optics to shape and polarize the laser beam, a Cs-Ne microcell where CPT interaction takes place and a photodiode to detect the transmitted light through the cell. The VCSEL laser is modulated with a 4.596~GHz from the local oscillator. A bias-tee is used to drive the laser both with a DC current ($I_{dc}$) and the 4.596~GHz signal for modulation and optical sidebands generation. The output signal from the photodiode is used in two servo loops: laser frequency stabilization and local oscillator frequency stabilization. The local oscillator frequency is locked to the atomic transition frequency by correcting the VCPS bias voltage. The output 4.596~GHz from the HBAR-based frequency source is mixed with the 4.596~GHz signal from a reference microwave synthesizer (RS SMB100A) driven by a hydrogen maser. The 190~kHz beatnote is filtered and counted with a frequency counter. A PC, not shown here, allows to compute and evaluate Allan deviation from the counter data. CL: collimation lens, NDF: neutral density filter, $\lambda$/4: quarter-wave plate.}
\label{fig:cpt-exp}
\end{figure}

Figure \ref{fig:cpt-exp} presents the Cs CPT clock experimental set-up. The laser source is a 25 MHz-linewidth custom-designed VCSEL tuned at 894.6 nm on the Cs D$_1$ line \cite{Al-Samaneh:APL:2012, Gruet:OE:2013}. The laser injection current is directly modulated at 4.596~GHz by a local oscillator
to generate two phase-coherent first-order optical sidebands frequency-split by 9.192~GHz for CPT interaction. Two different microwave sources were tested for comparison. The first source is the HBAR-based frequency source described in this article. The second source is a commercially-available microwave frequency synthesizer (Rohde-Schwarz RS SMB100A) driven by a high-stability 10 MHz quartz oscillator. The output laser beam is collimated with a collimation lens to reach a beam diameter of 2~mm, attenuated in power with a neutral density filter to a total laser power of about 30 $\mu$W and circularly polarized thanks to a quarter-wave plate. The bi-chromatic optical field interacts with Cs atoms confined in a micro-fabricated vapor cell whose architecture is described in \cite{Hasegawa:SA:2011}. A Ne buffer gas with total pressure of about 113 Torr (at 84$^{\circ}$C) is used to increase the time for Cs atoms to reach the cell walls. In this so-called Dicke regime \cite{Dicke}, the Doppler effect is canceled and a narrow microwave CPT clock resonance of a few kHz can be detected. The presence of Ne buffer gas shifts the Cs atom clock frequency by about $+78$~kHz (referred to the unperturbed Cs atom frequency = 9.192 631 770~GHz). In the optical domain,
the presence of buffer gas causes a broadening of optical transitions of about 10.85 MHz/Torr \cite{Pitz:PRA:2009}. The microcell temperature is stabilized to within 1 m$^\circ$C around 84$^{\circ}$C where the CPT signal height is maximized. The Cs-Ne cell is surrounded by a solenoid applying a static magnetic field of 10 $\mu$T in order to raise the Zeeman degeneracy and to isolate the hyperfine clock transition $|F = 3, m_F = 0\rangle \rightarrow |F = 4, m_F = 0\rangle$. The ensemble is inserted into a cylindrical mu-metal magnetic shield in order to prevent magnetic perturbations from the environment. The laser power transmitted through the cell is detected by a photodiode. The output signal of the photodiode is used in two main servo loops. The first servo loop aims to stabilize the laser frequency on the position of maximum optical absorption. For this purpose, the laser DC current is modulated (modulation frequency FM of 49~kHz) and the signal at the output of the photodiode is synchronously demodulated with the lockin-amplifier LA1 that generates a zero-crossing error signal. The latter is processed in a PI controller and used to correct with a bandwidth of a few kHz the laser current for stabilization of the laser frequency. The second servo loop is used to stabilize the local oscillator frequency onto the CPT clock resonance. Using the HBAR-based frequency source, the experimental procedure is as follows. First, the HBAR temperature is adjusted to about 68$^{\circ}$C to tune roughly the local oscillator frequency to the clock frequency. Secondly, the bias voltage of the VCPS is slowly swept with a ramp voltage to scan and detect the CPT resonance. Once the CPT resonance is detected, the bias voltage of the VCPS is sinusoidally modulated (FM = 932 Hz and modulation depth of 2.5~kHz) to modulate the LO frequency and the signal at the output of the photodiode is synchronously demodulated with the lockin-amplifier LA2. The zero-crossing error signal at the output of the lockin amplifier LA2 is processed in a PI controller and fed back to the VCPS bias voltage to stabilize the HBAR-source frequency to the CPT clock frequency.

The fractional frequency stability of the HBAR-based 4.596~GHz local oscillator is measured by beating it using a microwave mixer with the 4.596~GHz signal generated by a commercial microwave frequency synthesizer (Rohde-Schwarz RS SMB100A). The latter is driven by a reference hydrogen maser available in the laboratory. The 190~kHz beatnote is filtered using a 1.9 MHz low-pass filter, amplified with a low noise DC amplifier and counted by a frequency counter (HP53132A) referenced to the hydrogen maser. When using the 4.596~GHz RS SMB100A synthesizer as local oscillator of the clock, the correction signal is fed back to the 10 MHz quartz oscillator pilot. The output 10 MHz signal, locked to the atomic signal, is compared with a 10.190 MHz signal coming from a second commercial synthesizer driven by the hydrogen maser.

\subsection{Experimental results}

\begin{figure}[h!tb]
\centering
\includegraphics[width=\linewidth]{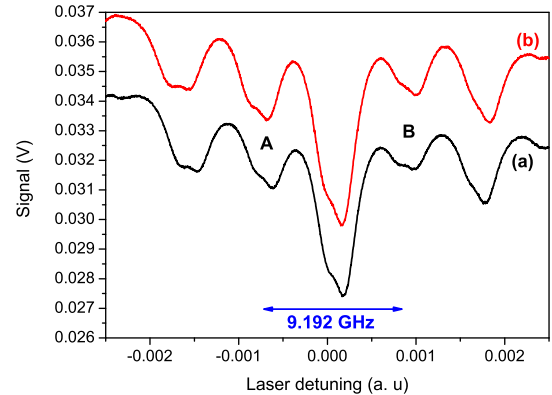}
\caption{Optical absorption lines detected by scanning the laser frequency. The VCSEL laser is modulated at 4.596~GHz. (a): with the HBAR-based frequency source, (b): with the laboratory-prototype microwave synthesizer (RS SMB100A).}
\label{fig:Abs}
\end{figure}

Figure \ref{fig:Abs} shows optical absorption lines detected at the output of the Cs-Ne cell by scanning the laser frequency. The laser frequency is modulated at 4.596~GHz by the local oscillator with a power of $+3$~dBm. Absorption lines detected with the HBAR-based frequency source (curve (a)) or the SMB100A frequency synthesizer (curve (b)) are compared.
When the 4.596~GHz signal is not applied, only both absorption doublets separated by 9.192~GHz (noted A and B on Fig. \ref{fig:Abs}) of the Cs D$_1$ line hyperfine structure are visible. Each absorption doublet is composed of two absorption peaks separated by 1.16~GHz (frequency separation between both energy levels of the Cs D$_1$ line excited state) that here are not well-resolved because of the buffer-gas induced optical broadening effect. When the 4.596~GHz modulation signal is applied, optical sidebands are generated and the effect of the 4.596~GHz modulation is clearly visible. In this case, both initial absorption doublets combine into a central doublet with amplitude that increases with increased microwave power. At the end, the microwave power is fixed at $+3$~dBm where the CPT signal is maximized. Similar absorption spectra are obtained with the HBAR-based frequency source or the high-performance commercial microwave synthesizer for same microwave power.

\begin{figure}[h!tb]
\centering
\includegraphics[width=\linewidth]{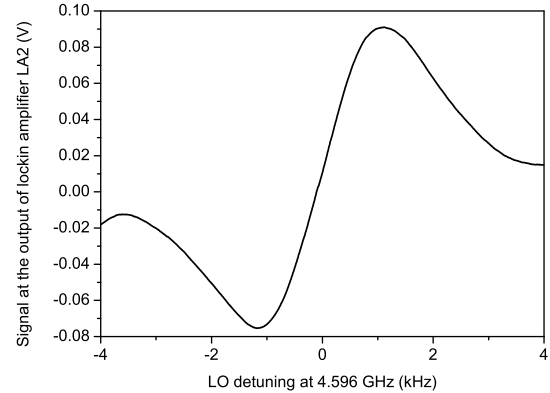}
\caption{CPT resonance (output of the lockin amplifier LA2) detected in the Cs-Ne microcell using the 4.596~GHz HBAR-based local oscillator.}
\label{fig:cpt}
\end{figure}

Figure \ref{fig:cpt} displays the CPT clock signal detected at the direct output of the lockin amplifier LA2 when the HBAR-based source frequency is slowly scanned around 4.596 354~GHz. 
We extract from this figure that the CPT resonance linewidth $\Delta \nu$ at 9.192~GHz is about 7~kHz. The frequency discriminator slope at the output of the lockin-amplifier is $8.2\times 10^{-5}$ V/Hz.

\begin{figure}[h!tb]
\begin{center}
\includegraphics[width=0.85\linewidth]{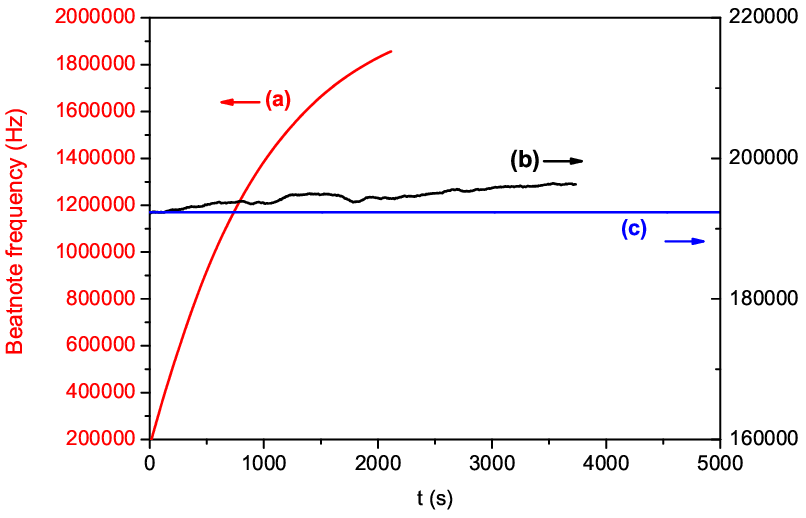} \\
(a) \\
\includegraphics[width=0.85\linewidth]{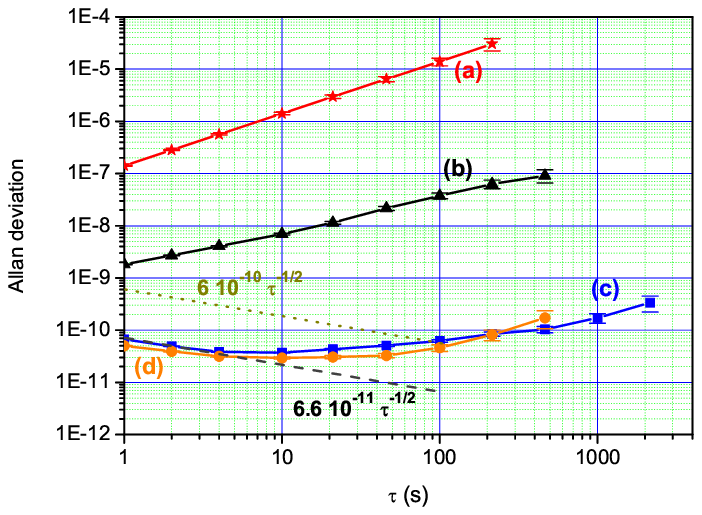} \\
(b)
\end{center}
\caption{Local oscillator frequency versus time (a) and corresponding Allan deviation (b). Different configurations were tested. (a): Free-running HBAR-based source without HBAR temperature control, (b): Free-running HBAR-based source with HBAR temperature control, (c): HBAR-based source with HBAR temperature control and locked to the atomic transition frequency, (d): laboratory-prototype microwave synthesizer (RS SMB100A) source locked to the atomic transition frequency. The HBAR is stabilized at 68$^{\circ}$C. The Cs-Ne microcell temperature is 84$^{\circ}$C. The total laser power incident in the cell is 30 $\mu$W. The dotted line, with a slope of 6 $\times$ 10$^{-10}$ $\tau^{-1/2}$, shows typical short-term frequency stability specifications of MACs. The dashed line, with a slope of 6.6 $\times$ 10$^{-11}$ $\tau^{-1/2}$, is a fit for short integration times to the Allan deviation obtained with the HBAR-based local oscillator.}
\label{fig:allandev} 
\end{figure}

Figures \ref{fig:allandev} (a) and (b) show respectively the local oscillator frequency versus time and corresponding Allan deviation versus the averaging time $\tau$. Different configurations were tested: free-running HBAR-based local oscillator without temperature control of the HBAR resonator (a), free-running HBAR-based local oscillator with temperature control of the HBAR resonator (b), HBAR-based local oscillator with HBAR temperature control and locked to the atoms (c), SMB100A synthesizer-based local oscillator locked to the atomic transition (d). Without HBAR temperature stabilization, the fractional frequency stability of the free-running HBAR-based source is measured to be $1.4 \times 10^{-7}$ $\tau$, that corresponds to a typical variation of the oscillator frequency of 643 Hz/s. With HBAR temperature control, the Allan deviation of the free-running HBAR-based source is improved at the level of $1.8\times 10^{-9}$ at 1~s and $3.7\times 10^{-8}$ at 100~s, demonstrating an improvement factor of about 670 at $\tau$ = 100~s compared to the non-temperature-stabilized regime. When locked to the atomic transition frequency, the HBAR-source exhibits a fractional frequency stability of $6.6\times 10^{-11}$ $\tau^{-1/2}$ up to 5~s, limited by the signal-to-noise ratio of the CPT signal. This measured short-term frequency stability, better than those reported in \cite{Yu:UFFC:2009, Boudot:BAW:2011}, is for $\tau$ up to 100~s well below the typical short-term stability specification of $6\times 10^{-10}$ $\tau^{-1/2}$ required in miniature atomic clocks. After 5~s, the clock frequency stability was found to be limited by laser power effects. At 100~s, the stabilization of the HBAR-oscillator to the atomic transition frequency allows to improve further its frequency stability by about 600 to reach the level of $6.8\times 10^{-11}$. For comparison, a frequency stability test was performed by using as local oscillator a laboratory-prototype microwave synthesizer (RS SMB100A) driven by a 10 MHz quartz oscillator. It is observed in this case that the short-term fractional frequency stability of the clock is recorded at the level of 5 $\times$ 10$^{-11}$ $\tau^{-1/2}$ up 5~s. Nevertheless, the observed difference between the two stability curves (HBAR-based source and laboratory-type synthesizer) are within the measurement errors and can be attributed to different loop settings. These results show that the performance of the HBAR-based oscillator is well suited for realizing miniature atomic clocks with excellent frequency stabilities.

\section{Discussions}
\label{sec:disc}

The work presented in this article is an original approach for the development of HBAR-sources dedicated to be used as local oscillators in MACs. Initially, tuning finely the output frequency of a HBAR resonator is not obvious due to the difficulty to control accurately thicknesses of materials during the fabrication process. Additionally, HBARs currently exhibit a high temperature coefficient of frequency of 10 to 20 ppm/$^\circ$C. Techniques to reduce the HBAR temperature sensitivity exist \cite{Yu:UFFC:2007} but often at the expense of a degradation of the Q-factor and supplemental technological steps bringing increased complexity. Eventually, an inherent characteristic of HBAR is their multi-mode spectrum that complicates their use in an oscillator and forces one to filter the response to select a single resonance. In this article, we tried to convert these drawbacks into strong key points to adapt the HBAR to MAC applications. The high-temperature sensitivity is an advantage for coarse tuning of the HBAR frequency and hence
compensate for unreachable thickness resolutions for reaching the Cs transition frequency.
Moreover, the fine tuning of the HBAR oscillator frequency with a voltage-controlled phase shifter in the loop prevents the use of a high-power consuming direct digital synthesizer. In our system, the voltage-to-frequency tuning with the VCPS was about 8~kHz/V. Then, a supply voltage with a resolution of 1 $\mu$V allows a satisfying frequency resolution of 8 mHz.

While no integration and packaging efforts were done in this work, related activities demonstrate that this strategy is compatible with low volume and low power consumption requirements for MAC applications.
In terms of compacity, the dimensions of the HBAR are driven by the electrode area and the total material stack thickness. The acoustic energy confinement in the substrate requires electrodes dimensions ``very'' large with respect to the acoustic wavelength. Operating above 2~GHz induces wavelengths lower than 6~$\mu$m and a 1$\times$1~mm large HBAR transducer meets the acoustic energy confinement conditions. Hence, HBAR meets clearly the compact resonator requirement. The dimensions of the oscillator circuit have not been considered in this investigation but have been demonstrated at the integrated chip level in the literature \cite{hajimiri1999design}.

	The power consumption of the HBAR-local oscillator is mainly driven by the microwave sustaining amplifier and the resonator thermal control. The power consumption of the VCPS (capacitive load polarized with a DC voltage) is negligible. In our system, the main contribution to the consumption budget is the HBAR temperature control circuit with over 900~mW needed to heat the packaged HBAR. Such a power consumption is not acceptable for embedded applications. The best way to reduce such a large current would be to confine the heating to the bare resonator rather than the whole packaged device. Ideally, the temperature set point of the HBAR should be fixed at about 85 -- 90$^{\circ}$C to support a MAC operation typical temperature range of $-20$ to $+80^{\circ}$C without using a high-power consuming Peltier element. MEMS metal oxide gas sensors, which require operating temperatures several hundreds of celsius degrees above room temperature, have met this challenge by fitting the sensing element to heat insulating hinges, patterning the sensing element with the heating coil, and confining the heating to the part of interest, in our case the volume in which the acoustic energy is stored, reaching sub-100~mW consumption to reach 300$^{\circ}$C for 1000~$\mu$m$^2$ large devices \cite{briand2013micromachined}. 
We think that the power consumption of a HBAR oscillator for MAC applications could be reduced at a maximum of 10--20~mW.
Nevertheless, a question remains. In standard MACs, the output useful signal frequency for end-users is 10 MHz. In a potential HBAR-based oscillator Cs vapor cell MAC, the LO is frequency-stabilized at 4.596~GHz. This microwave frequency is not well-adapted for standard widespread applications. This issue would impose to use a frequency divider to downconvert the 4.596~GHz signal to about 100 or 10~MHz, adding a power consumption of 10--20~mW if dedicated components are used \cite{Razavi, Haijun, Felder}.

	We demonstrated that HBAR oscillators exhibit ultra-low phase noise performances. Hence, it is interesting to note that MACs using HBAR-based oscillators could allow the development of time-frequency references combining in a single device excellent phase noise and long-term frequency stability properties, opening potentially the MAC technology to a wider spectrum of applications. Up to date, the short-term fractional frequency stability of MACs is mainly limited by the laser FM noise and not by the LO intermodulation effect. Thus, to date, improving the LO noise with HBAR-based oscillators should not improve the clock short-term stability. However, we demonstrated that the use of a HBAR-LO rejects the Dick effect contribution to a level close to the clock shot noise limit. This aspect could be of great interest in the future if new-generation high-performance MACs using miniature low-consumption lasers with ultra-narrow spectral linewidth \cite{Sopra} and reduced AM noise were developed.
	%
	\section{Conclusions}\label{sec:conclu}

	We reported the design strategy and the use of a double-port AlN/Sapphire 2.298~GHz HBAR-oscillator-based 4.596~GHz frequency synthesizer devoted to be used as a local oscillator in a Cs vapor cell CPT-based atomic clock.
	Dedicated techniques, exploiting the HBAR temperature sensitivity for coarse frequency tuning and the use of a VCPS in the oscillator loop, were proposed to ensure the HBAR-oscillator frequency to be resonant with the half of Cs atom clock frequency (4.596~GHz).
	The HBAR presents a Q-factor of about 24000 and a temperature sensitivity of $-$23 ppm/$^\circ$C. The 4.596~GHz output signal exhibits exceptional phase noise performances of $-105$~dBrad$^2$/Hz at 1~kHz offset frequency, rejecting
	the intermodulation effect contribution to the clock short-term fractional frequency stability at the level of $4.3\times 10^{-13}$, a value comparable to the clock quantum noise limit. A Cs microcell-based CPT clock was tested using the HBAR-source, demonstrating a clock short-term
frequency stability of $6.6\times 10^{-11}$, limited by the CPT resonance signal-to-noise ratio and not by the LO phase noise. Discussions were reported to evaluate the potential of this technology to be implemented
in low-volume and low power consumption miniature atomic clocks.

\section*{Acknowledgments}\label{sec:acknow}
This work has been funded by LabeX FIRST-TF. The HBAR was manufactured by CEA/LETI in the framework
of the ORAGE RAPID project funded by the French Direction G\'en\'erale de l'Armement (DGA) agency. 


%
\end{document}